\documentclass[10pt,letterpaper]{article}

\usepackage{amsmath,amssymb,graphicx}
 \usepackage{bm}
\usepackage{opex3}
 \usepackage[dvipsnames]{xcolor}
\colorlet{LightRubineRed}{RubineRed!70!}
\colorlet{Mycolor1}{green!10!orange!90!}
\colorlet{Mycolor2}{green!30!orange!60!red!40!}
\definecolor{lgreen}{RGB}{118,238,0}
\definecolor{green}{RGB}{0,190,0}
\definecolor{brown}{RGB}{205,100,0}
\definecolor{yellow}{RGB}{200,125,150}
\definecolor{RubineRed}{RGB}{206,0,88}
\definecolor{brown(traditional)}{rgb}{0.59, 0.29, 0.0}
\definecolor{bistre}{rgb}{0.24, 0.17, 0.12}
\definecolor{auburn}{rgb}{0.43, 0.21, 0.1}
\definecolor{armygreen}{rgb}{0.29, 0.33, 0.13}
\definecolor{bazaar}{rgb}{0.6, 0.47, 0.48}
\definecolor{darkolivegreen}{rgb}{0.33, 0.42, 0.18}
\definecolor{burgundy}{rgb}{0.5, 0.0, 0.13}
\definecolor{amaranth}{rgb}{0.9, 0.17, 0.31}
\definecolor{azure(colorwheel)}{rgb}{0.0, 0.5, 1.0}
\definecolor{ballblue}{rgb}{0.13, 0.67, 0.8}

\begin{document}

%

\title{Electromagnetic cloaking in convex and concave media with surface modelled as a parameterised function}

\author{H. ~H. ~Sidhwa$^{1,*}$ R.~P.~Aiyar$^{2}$ and S.~V.~Kulkarni$^1$ }

\address{$^1$Department of Electrical Engineering, Indian Institute of Technology Bombay, Mumbai 400076, India\\
$^2$Centre for Research in Nanotechnology and Science, Indian Institute of Technology Bombay, Mumbai 400076, India}

\email{$^*$hoshedar@iitb.ac.in} 



\begin{abstract}
The onset of transformation optics has opened avenues for designing of a plenitude of applications related to propagation of electromagnetic 
waves in anisotropic media. In this paper, an algorithm is proposed using a coordinate transformation and a piecewise function for the 
purpose of designing a three dimensional cloak having an arbitrary geometry
which could be convex or non-convex in nature. {\color{black}The surfaces of the cloak as well as of the body under consideration are assumed to be conformal to each other}. For an arbitrary geometry, the coordinate system needed to 
model the surface can be a non-orthogonal system.
For the purpose of verification of the algorithm, a ray tracing process is carried out for an ellipsoid as well as for a concave surface having axial symmetry.
In order to solve the Hamiltonian equation for the purpose of ray tracing, {\color{black}the process of finding the derivatives analytically,
for an arbitrary geometry as considered here, becomes very cumbersome. Here, a numerical method is described which provides a better approximation
to the partial derivatives than the conventional finite difference approach based on forward differences}.
\end{abstract}


\section{Introduction}


The development in the fields of metamaterials and transformation optics in the past decade has enabled scientists to envision and realise
devices/applications such as {\color{black}electromagnetic cloaks \cite{geometry, geometry_rel, cummer, pendry_first}, perfect lenses \cite{pendry_lens} and spacetime cloaks \cite{spacetime}}.
{\color{black}A formalism  of transformation optics which also encompasses gravitational curvature has been illustrated in \cite{thompson}.
A  technique which uses transformation between spacetime manifolds with application to cloaking is shown in \cite{thompson2}}.
\textcolor{black}{Cloaks with non-conformal inner and outer boundaries have been analysed at by Li and Li \cite{nonconfor}}. The concept of cloaking or invisibility which is one
such application, can be explained as bending the path of a ray around the body of interest so as to give an impression that the
ray is travelling in a straight line uninterrupted by any obstacle \cite{hoshi}. In  2006, Pendry, et al. \cite{pendry_first} and Leonhardt \cite{leo} carried out 
 a coordinate transformation in order to alter the material characteristics, which  would in turn cause a change in the formulation 
of Maxwell's equations. The idea of ray tracing in the transformed media was used by Pendry, et al. \cite{pendry} to verify the concept for spherical and cylindrical cloaks using Cartesian  tensors.
A generalised method for designing arbitrarily shaped cloaks using the approach of coordinate transformation has been discussed by Li and Li in \cite{chen}.
In \cite{hoshi1}, an algorithm  is proposed for tracing the path of a ray in an arbitrarily shaped cloak in three dimensions using spherical polar coordinates ($\theta, \phi $), which requires that
the contour of the outer boundary of the cloak should be a single valued function of ($\theta, \phi $).
\textcolor{black}{In this paper, an algorithm is propounded for cloaking of an arbitrarily shaped body in three dimensions using a parametric formulation
without the above constraint of the need of a single valued function.} \textcolor{black}{The requirement of a numerical method to calculate the derivatives of the 
Hamiltonian stems from the limitation that when the geometry under consideration is complex, analytical solutions for transformation 
equations may not exist}. \textcolor{black}{The body under consideration could either have a concave or convex 
geometry. The surfaces of the cloak as well as of the body under consideration are assumed to be conformal to each other. This representation can either be a piecewise continuous  polynomial similar to Finite Element Method \cite{jing} or
be defined in terms of Non-Uniform Rational B-Spline (NURBS) \cite{nurbs}. For the purpose of illustrating the technique, an axisymmetric
body is represented using a piecewise linear function. That enables the algorithm to cloak a concave shaped body.
Higher order functions can also be used since the procedure remains unchanged. In numerical differentiation,  truncation and round-off errors 
cause numerical instability while using the conventional finite difference approximation based on forward differences \cite{numerical_burden}}.
For the purpose of ray tracing, a Hamiltonian is calculated using a technique which has been adapted  from that reported in \cite{pendry}. 
\section{Mathematical formulation}
\label{sec:formulation}

\begin{figure}
\centerline{\includegraphics[width=9 cm]{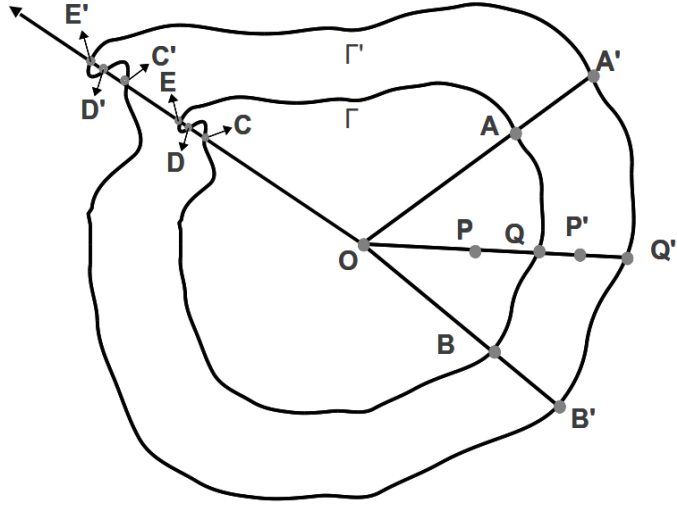}}
\caption{Coordinate transformation for an arbitrary geometry.}
\label{fig:coodtrx}
\end{figure}

{\color{black}In order to cloak an arbitrarily shaped object, it is coated with a thin layer of a material such that any electromagnetic
wave incident on the body will move around the object with zero reflection. For practical reason, the coating thickness is uniform, hence it can be considered}
{\color{black}that the cloak would have the same topology as the object to be cloaked }. Then, the formulation can be stated as follows:

Consider $O$ to be the centre of the concentric bodies as shown in (Fig.~\ref{fig:coodtrx}). 


Consider three points $A, B$ and $Q$ on $\Gamma$. A surface $\Gamma'$ is obtained by scaling $\Gamma$ by a suitable factor.

In the figure, the scaling implies:
\begin{eqnarray*}
\dfrac{OA}{OA'}=\dfrac{OB}{OB'}=\dfrac{OQ}{OQ'}=\tau.   
\end{eqnarray*}

Here $\tau<1$ is the measure of the thickness of the cloak. Any surface can be represented by an implicit equation $f(x,y,z)=0$.
A parametric representation, which need not be unique, of the same surface can be written in terms of two parameters:
\begin{eqnarray*}
 \bm{r}=\left(x(u_2,u_3),\,y(u_2,u_3),\,z(u_2,u_3)\right)=\bm{R}(u_2,u_3).
\end{eqnarray*}

The outer surface of the cloak is described using parametric coordinates as:

\begin{eqnarray}
{\color{black}
 \bm{r}=\bm{R}(u_2, u_3).}
\end{eqnarray}
The magnitude of the position vector $\bm{r}$ will be $R(u_2, u_3)$ for any point on the surface.
In \cite{hoshi1}, a parametric representation has been used in terms of spherical polar coordinates
$(\theta,\phi)$. Such an approach has a limitation in modelling general surfaces that can be concave because 
$(\theta,\phi)$ representation requires $\bm{R}$ to be a single valued function of $(\theta,\phi)$.
The outer surface of the cloak in (Fig.~\ref{fig:coodtrx}) cannot be modelled by the algorithm mentioned in 
\cite{hoshi1} since points $C,D$ and $E$ lie along the same radial path, hence for the same values of ($\theta, \phi$), $\bm{R}(\theta,\phi)$ will
have different values.


 Other
parametric representations like NURBS or planar polynomial parametric approximation do not have such limitations.
Here we describe a formulation in terms of an arbitrary parametric representation $(u_2,u_3)$. The following coordinate transformation is applied.


\begin{eqnarray}\label{eq: transform}
 r'&=\left(1-\tau\right)r+\tau R.
 \end{eqnarray}
 Here, $r$ is the coordinate of any point $P$ inside the volume enclosed by $\Gamma'$. Owing to this transformation,
 any point $P$ within the outer surface $\Gamma'$ is transformed to $P'$ in the annular space between $\Gamma$ and $\Gamma'$. The unprimed 
 coordinates apply to the whole domain while the primed coordinates apply to the annular region. A point 
 transforms from $P(x,y,z)$ in the original system to $P'(x',y',z')$ in the primed system.\\
 Due to transformation invariance, the unit vectors in the original system and the transformed system must be equal.
\begin{eqnarray}
 \dfrac{x'}{r'}=\dfrac{x}{r}, \hspace{10 mm} \dfrac{y'}{r'}=\dfrac{y}{r}  ,  \hspace{10 mm} \dfrac{z'}{r'}=\dfrac{z}{r}.
\end{eqnarray}
Expressing the equation in the tensor notation and using Eq. (\ref{eq: transform}),
\begin{eqnarray}
 x'_i &=\dfrac{r'}{r} x_i \nonumber.
\end{eqnarray}
Here the subscripts vary from 1 to 3 and we shall use Einstein summation convention.\\
Any point inside the cloak can be described in terms of three variables $(r, u_2, u_3)$:
\begin{eqnarray}\label{eq: jacob1}
{\color{black}
 x_i=r\dfrac{{R}_i(u_2, u_3)}{R}=r\hat{{R}}_i(u_2, u_3)}.
\end{eqnarray}
The unit vector $\hat{\bm{R}}$ can be expressed in terms of its components as
\begin{align}
\hat{\bm{R}}= \left[ {\begin{array}{ccccc}
   \hat R_1\\
    \hat R_2\\
    \hat R_3
  \end{array} } \right]
\end{align}

In terms of the primed variables, the point can denoted by:

\begin{align}
\color{black}
x'_i &=\color{black} r'\dfrac{{R}_i(u'_2, u'_3)}{R}=r'\hat{{R}}_i(u'_2, u'_3).
\end{align}
The Jacobian for the coordinate transformation can be written as:
 
\begin{eqnarray}
 \boldsymbol{\Lambda}= \left[ 
  \begin{array}{ccc}
 \dfrac{\partial x'}{\partial x} & \dfrac{\partial x'}{\partial y} & \dfrac{\partial x'}{\partial z} \\
 \dfrac{\partial y'}{\partial x} & \dfrac{\partial y'}{\partial y} & \dfrac{\partial y'}{\partial z}\\
 \dfrac{\partial z'}{\partial x} & \dfrac{\partial z'}{\partial y} & \dfrac{\partial z'}{\partial z}
  \end{array}
  \right]
  \end{eqnarray}
In the tensor notation, it is expressed as:
 \begin{eqnarray}
 \Lambda^{i'}_l &=\dfrac{\partial x'_i}{\partial x_l}.
\end{eqnarray} 
The constitutive properties in the transformed space are:
\begin{subequations}
\begin{eqnarray}
 \boldsymbol{\epsilon}'&=\dfrac{\boldsymbol{\Lambda} \cdot \boldsymbol{\epsilon}\cdot \boldsymbol{\Lambda}^T}{|\boldsymbol\Lambda|}\\
 \boldsymbol{\mu}'&=\dfrac{\boldsymbol{\Lambda} \cdot \boldsymbol{\mu}\cdot \boldsymbol{\Lambda}^T}{|\boldsymbol\Lambda|}.
\end{eqnarray}
\end{subequations}
If the original space is assumed to be free space,
\begin{eqnarray*}
 \boldsymbol{\epsilon}=\bm{I}\epsilon_0 \hspace{25mm} 
 \boldsymbol{\mu}=\bm{I}\mu_0
\end{eqnarray*}
where $\bm{I}$ is the Identity tensor.
Relative permittivity and permeability in the transformed space can be expressed as:
\begin{eqnarray}\label{eq: epsilonprime}
 \boldsymbol{\epsilon}'=\boldsymbol{\mu'}=\dfrac{\bm{\Lambda} \cdot \bm{\Lambda}^T}{|\bm{\Lambda}|}.
\end{eqnarray}
The expression for Jacobian in the primed coordinates is:
%
 
 \begin{align}
 \Lambda_{il}&=\dfrac{\partial x'_i}{\partial x_l}=\dfrac{r'}{r}\left[\delta_{il}-\tau \dfrac{x'_i}{r'}\dfrac{x'_l}{r'} \dfrac{ R}{r'}+\tau\dfrac{r}{r'}\dfrac{x'_i}{r'} \dfrac{\partial R}{\partial x_l}\right] \nonumber \\
 \Lambda_{il}&=\dfrac{\partial x'_i}{\partial x_l}=\dfrac{(1-\tau)r'}{r'-\tau R}\left[\delta_{il}-\tau \dfrac{x'_i}{r'} \dfrac{ R}{r'}\dfrac{\partial r}{\partial x_l} 
 +\tau\dfrac{(r'-\tau R)}{(1-\tau)r'}\dfrac{x'_i}{r'} \dfrac{\partial R}{\partial x_l}\right]\nonumber \\
 \Lambda_{il}&=\dfrac{(1-\tau)r'}{r'-\tau R}\widetilde{\Lambda}_{il}.\label{eq: jacobprime}
 \end{align}

{\color{black}In Eq. (\ref{eq: jacob1}), the position vector is expressed in terms of variables $(r,u_2, u_3)$ which are considered over here  as the 
generalised curvilinear coordinates denoted by $(q^1,q^2,q^3)$ where}
\begin{align*}
 q^1=r \hspace{10mm} q^2=u_2 \hspace{10mm}  q^3=u_3.
\end{align*}
The  basis vectors $\bm{a}_i$ for $(i=1,2,3)$
are defined as follows:
\begin{eqnarray}
 \bm{a}_i=\dfrac{\partial \bm{r}}{\partial q^i}.
\end{eqnarray}
The differential position vector can be expressed in terms of Cartesian components:
\begin{align}\label{eq: dpos}
 d\bm{r}&=\sum_i  \bm{a}_i dq^i=\sum_j\hat{\bm{e}}_j dx_j=\sum_i\left[ \sum_j\hat{\bm{e}}_j\dfrac{\partial x_j}{\partial q^i}\right]dq^i \nonumber \\
\bm{a}_i&=\sum_j\hat{\bm{e}}_j \dfrac{\partial x_j}{\partial q^i}.
\end{align}
where $\hat{\bm {e}}_1,\; \hat{\bm {e}}_2,\; \hat{\bm {e}}_3$ are the standard Cartesian unit vectors.
This can be expressed as:

%

\renewcommand{\arraystretch}{2.5}
\[ \left[ \begin{array}{c}
\mathbf{a_1}  \\
\mathbf{a_2}\\
\mathbf{a_3}
\end{array} \right]=\left[ 
  \begin{array}{ccc}
  \dfrac{\partial x_1}{\partial q^1} & \dfrac{\partial x_2}{\partial q^1} & \dfrac{\partial x_3}{\partial q^1} \\
  \dfrac{\partial x_1}{\partial q^2} & \dfrac{\partial x_2}{\partial q^2} & \dfrac{\partial x_3}{\partial q^2}\\
  \dfrac{\partial x_1}{\partial q^3} & \dfrac{\partial x_2}{\partial q^3} & \dfrac{\partial x_3}{\partial q^3}
  \end{array}
  \right]
\left[ \begin{array}{c}
\hat{\mathbf{e}}_1 \\
\hat{\mathbf{e}}_2 \\
\hat{\mathbf{e}}_3
\end{array} \right]
\]

Since $q^1=r$, we can write using Eq. (\ref{eq: jacob1}):

\renewcommand{\arraystretch}{2.5}
\begin{eqnarray}
 J&= \left[ 
  \begin{array}{ccc}
  \dfrac{\partial x_1}{\partial q^1} & \dfrac{\partial x_2}{\partial q^1} & \dfrac{\partial x_3}{\partial q^1} \\
  \dfrac{\partial x_1}{\partial q^2} & \dfrac{\partial x_2}{\partial q^2} & \dfrac{\partial x_3}{\partial q^2}\\
  \dfrac{\partial x_1}{\partial q^3} & \dfrac{\partial x_2}{\partial q^3} & \dfrac{\partial x_3}{\partial q^3}
  \end{array}
  \right]\nonumber =\left[ 
%
\begin{array}{ccc}
  \hat{{R}}_1 &  \hat{{R}}_2 &  \hat{{R}}_3 \\
  r\dfrac{\partial  \hat{{R}}_1}{\partial q^2} & r\dfrac{\partial \hat{{R}}_2}{\partial q^2} & r\dfrac{\partial  \hat{{R}}_3}{\partial q^2}\\
  r\dfrac{\partial  \hat{{R}}_1}{\partial q^3} & r\dfrac{\partial \hat{{R}}_2}{\partial q^3} & r\dfrac{\partial  \hat{{R}}_3}{\partial q^3}\\

  \end{array}
  \right]
  \end{eqnarray}

In order to define the gradient in terms of generalised coordinates, a differential volume element $dV$ is required.
\begin{eqnarray*}
 dV=\bm{a}_1\cdot \bm{a}_2\times \bm{a}_3\; dq^1\; dq^2\;dq^3= \Omega \; dq^1\; dq^2\;dq^3.
\end{eqnarray*}
The gradient of a scalar function $F$ can be expressed as:
 
 \begin{eqnarray}
 \nabla F\left(q^1,~ q^2, ~q^3\right)&=\dfrac{1}{\Omega}\left[(\bm{a}_2\times \bm{a}_3) \dfrac{\partial F}{\partial q^1}+(\bm{a}_3\times \bm{a}_1)\dfrac{\partial F}{\partial q^2}
 +(\bm{a}_1\times \bm{a}_2) \dfrac{\partial F}{\partial q^3}\right]
 \end{eqnarray}
The Jacobian, Eq. (\ref{eq: jacobprime}) can now be expressed as

\begin{eqnarray}\label{eq: jacobfinal}
\color{black}
   \widetilde{\Lambda}_{il} &=\color{black} \left[\delta_{il}-\tau \dfrac{x'_i}{r'} \dfrac{ R}{r'}\dfrac{(\bm{a}_2\times \bm{a}_3)_l}{\Omega}+\tau\dfrac{(r'-\tau R)}{(1-\tau)r'}\dfrac{x'_i}{r'} \dfrac{\partial R}{\partial q^2}\dfrac{(\bm{a}_3\times \bm{a}_1)_l}{\Omega}\right.\nonumber \\
   &\quad  \left.\color{black}+\tau\dfrac{(r'-\tau R)}{(1-\tau)r'}\dfrac{x'_i}{r'} \dfrac{\partial R}{\partial q^3}\dfrac{(\bm{a}_1\times \bm{a}_2)_l}{\Omega}\right] 
\end{eqnarray}

{\color{black}Using \cite{thompson2} where transformation optics is expressed as a geometric formulation on spacetime manifolds, an expression similar
to Eq. (\ref{eq: jacobfinal}) has been generated}.\\
{\color{black}The original coordinate system $(q^1, q^2, q^3)$ is transformed to the primed coordinate system $(q'^1, q'^2, q'^3)$ such that
$(q^1=r)\rightarrow (q'^1=r')$ whereas both $ q^2 ~\&~ q^3$ remain unchanged in the primed coordinate system}.

\section{Axisymmetric cloak}
 The special case of an axisymmetric object is considered.
 \begin{align}
 x_1 =r \hat{{R}}_1(q^2) cos \phi \hspace{10mm} x_2 =r \hat{{R}}_1(q^2) sin\phi \hspace{10mm}
  x_3 = r \hat{{R}}_2 (q^2)
 \end{align}
 
 The object is modelled in the $XZ$ plane as a curve with a single parameter $q^2$. A three dimensional object is 
 generated by rotating the curve about the $Z$ axis by $2 \pi$.
 
 \renewcommand{\arraystretch}{2.5}
\begin{eqnarray}
J=\left[ 
  \begin{array}{ccc}
  \hat{{R}}_1 cos\phi &  \hat{{R}}_1 sin\phi &  \hat{{R}}_2 \\
  r\dfrac{\partial  \hat{{R}}_1}{\partial q^2}cos\phi & r\dfrac{\partial \hat{{R}}_1}{\partial q^2}sin\phi & r\dfrac{\partial  \hat{{R}}_2}{\partial q^2}\\
  -r\hat{{R}}_1sin\phi & r\hat{{R}}_1 cos\phi & 0\\
  \end{array}
  \right]
  \end{eqnarray}
The  basis vectors can be written using Eq. (\ref{eq: dpos}):
 
 \begin{eqnarray} \label{eq: covariantbase}
 \left[ \begin{array}{c}
\mathbf{a_1}  \\
\mathbf{a_2}\\
\mathbf{a_3}
\end{array} \right]=\left[ 
 \begin{array}{ccc}
  \hat{{R}}_1 cos\phi   &  \hat{{R}}_1sin\phi  &  \hat{{R}}_2  \\
  r\dfrac{\partial  \hat{{R}}_1}{\partial q^2}cos\phi   & r\dfrac{\partial \hat{{R}}_1}{\partial q^2}sin\phi & r\dfrac{\partial  \hat{{R}}_2}{\partial q^2} \\
  -r\hat{{R}}_1 sin\phi   & r\hat{{R}}_1 cos\phi  & 0\\
  \end{array}
  \right]
\left[ \begin{array}{c}
\hat{\mathbf{e}}_1 \\
\hat{\mathbf{e}}_2 \\
\hat{\mathbf{e}}_3
\end{array} \right]
\end{eqnarray}
The  Jacobian now becomes:
 
 \begin{eqnarray}\label{eq: jacob2}
   \widetilde{\Lambda}_{il} &= \left[\delta_{il}-\tau \dfrac{x'_i}{r'} \dfrac{ R}{r'}\dfrac{(\bm{a}_2\times \bm{a}_3)_l}{\Omega}
  +\tau\dfrac{(r'-\tau R)}{(1-\tau)r'}\dfrac{x'_i}{r'} \dfrac{\partial R}{\partial q^2}\dfrac{(\bm{a}_3\times \bm{a}_1)_l}{\Omega}\right]
  \end{eqnarray}

\subsection{Hamiltonian}\label{hamil_sec}
Since there is no loss of energy while the wave propagates through the medium,  
the Hamiltonian can be found in order to trace the path of the wave in the cloaked medium \cite{pendry}.

\begin{eqnarray}\label{eq: hamil1}
 {H}&=\left<\bm{k}\mid \boldsymbol{\epsilon}'\mid \bm{k}\right>-|\boldsymbol{\epsilon}'|.
\end{eqnarray}
Here $\bm{k}$ is the propagation vector in the cloak.

The first and second terms of Eq. (\ref{eq: hamil1}) become:
\begin{eqnarray*}
 {H_1}&=\dfrac{\left<\bm{k}|\bm{\Lambda} \cdot \bm{\Lambda}^T|\bm{k}\right>}{|\bm{\Lambda}|} \\
 {H_2}&=\dfrac{|\bm{\Lambda}| \cdot |\bm{\Lambda}^T|}{|\bm{\Lambda}|^3}=\dfrac{1}{|\bm{\Lambda}|}.
\end{eqnarray*}
The Hamiltonian equation is:
\begin{align}\label{eq: hamil2}
 H&=\left<\bm{k}|\bm{\Lambda} \cdot \bm{\Lambda}^T|\bm{k}\right>-1\\ \nonumber
 &=\left<\bm{k}|\widetilde{\bm{\Lambda}} \cdot \widetilde{\bm{\Lambda}}^T|\bm{k}\right>
 -\left[ \dfrac{r'-\tau R}{(1-\tau)r'}\right]^2.
\end{align}
When Eq. (\ref{eq: jacob2}) is substituted in Eq. (\ref{eq: hamil2}), it is observed that the Hamiltonian H in Eq. (\ref{eq: hamil2}) is 
a function of generalised coordinates $(q^1, q^2, q^3)$ and propagation vector $\bm k$.

For the purpose of ray tracing, the path can be parameterised using the Hamiltonian in Eq. (\ref{eq: hamil2}) as explained in \cite{orlov}:
\begin{eqnarray*}
\dfrac{d\bm{x}}{dt}&=\dfrac{\partial H}{\partial\bm{k}} \\
\dfrac{d\bm{k}}{dt}&=-\dfrac{\partial H}{\partial \bm{x}}
\end{eqnarray*}
where $t$ is the parameterisation  variable and $\bm{x}$ is the position vector. Ray tracing is carried out by
solving $\left(\bm{x}, \bm{k}\right)$ as a function of $t$ starting from $t=0$, which corresponds to the point
at which the incident wave touches the cloak.

In  the technique proposed by Pendry, et al. \cite{pendry}, the independent variables for the Hamiltonian are
in terms of spherical polar coordinates ($r, \theta, \phi$), whereas the ray tracing is done in terms of Cartesian coordinates.
At every step in the algorithm, an inverse transformation from Cartesian to the spherical polar coordinate system is applied, which is trivial in that case.
But in the coordinate system used in this paper, there is no explicit relation for inverse transformation from
($x,y,z$)  to $(q^1, q^2, q^3)$ coordinate system. That in turn implies that the ray tracing algorithm also needs to be
written in terms of the generalised coordinates.

\begin{eqnarray*}
\dfrac{\partial H}{\partial{k_i}}&=\dfrac{d{x'_i}}{dt} =2\left[\widetilde{\Lambda}\widetilde{\Lambda}^T\right]_{jl} k_l
\end{eqnarray*}
where $\bm{k}$ is still in the Cartesian system.


In order to solve the coupled differential equations, Eqs. (\ref{eq: contravariant_array})   
and Eq. (\ref{eq: contravariant3}) mentioned later, the derivative of the Hamiltonian is required to be evaluated with respect to the generalised
coordinates $(q'^{(1)},q'^{(2)},q'^{(3)})$, see Eq. (\ref{eq: jacobfinal}). {\color{black}The analytical expression for that would be too tedious to evaluate}.
 We describe a numerical method which
has better accuracy than the conventional finite difference approximation based on forward differences.
In the method we follow, we assume that the differential term $\Delta q'^{(i)} \hspace{2mm} i=1,2,3 $ is an infinitesimal variation in $q'^{(i)}$.
For the purpose of illustration, we assume that $i=3$ i.e. the numerical derivative with respect to $q'^{(3)}$ is being calculated in the following steps.

{\color{black} When $\Delta q'^{(3)}$ is real, the Taylor Series of the Hamiltonian H (for the first four terms neglecting higher order terms) can be expressed as}
\begin{eqnarray}\label{eq: taylor1}
\begin{split}
H(q'^{(1)},q'^{(2)},q'^{(3)}+\Delta q'^{(3)} )&= \color{black}\left[ 1+ \Delta q'^{(3)}\; \dfrac{\partial }{\partial q'^{(3)}}+ 
 \dfrac{1}{2!} \left( \Delta q'^{(3)}\; \dfrac{\partial }{\partial q'^{(3)}}\right)^2 \right.\\  &\quad \left.
 \color{black} +\dfrac{1}{3!} \left( \Delta q'^{(3)} \; \dfrac{\partial }{\partial q'^{(3)}}\right)^3
 \right]\color{black}H(q'^{(1)},q'^{(2)},q'^{(3)})
 \end{split}
\end{eqnarray}
{\color{black}If only the first order term is retained, the standard forward difference approximation for the derivative is obtained. The 
first neglected term is the $2^{nd}$ order term $(\Delta q'^{(3)})^2$. Instead, if it is assumed that $\Delta q'^{(3)}$ is imaginary, 
and is expressed as $j \Delta q'^{(3)}$, the Taylor series for $H(q'^{(1)},q'^{(2)},q'^{(3)} )$ can now be 
expressed in terms of $j \Delta q'^{(3)}$.}
\begin{align*}
H(q'^{(1)},q'^{(2)},q'^{(3)}+j\;\Delta q'^{(3)} )&=\color{black} H(q'^{(1)},q'^{(2)},q'^{(3)} )+ j\;\Delta q'^{(3)} \dfrac{\partial H }{\partial q'^{(3)}}\\
 & +\dfrac{1}{2!} \left( j\Delta q'^{(3)}\; \dfrac{\partial }{\partial q'^{(3)}}\right)^2 H 
 +\dfrac{1}{3!} \left( j\Delta q'^{(3)}\; \color{black}\dfrac{\partial }{\partial q'^{(3)}}\right)^3 H\\
 Im\left(H(q'^{(1)},q'^{(2)},q'^{(3)}+j\;\Delta q'^{(3)} )\right)&=\Delta q'^{(3)} \dfrac{\partial H }{\partial q'^{(3)}} - \dfrac{1}{3!} \left( \Delta q'^{(3)}\; \dfrac{\partial }{\partial q'^{(3)}}\right)^3 H.
 \end{align*}
 Thus,
 \begin{eqnarray}\label{eq: numerical1}
 \dfrac{\partial H}{\partial q'^{(3)}}=\dfrac{1}{\Delta q'^{(3)}}Im\left[H(q'^{(1)},q'^{(2)},q'^{(3)}+j \Delta q'^{(3)})\right].
\end{eqnarray}
{The order of the first neglected term  $(\Delta q'^{(3)})^3$ is three instead of two in the earlier case.
This technique described by Eq. (\ref{eq: numerical1}) is used for evaluating the partial derivative
numerically instead of the conventional finite difference method. This method does not suffer from round-off errors
if the derivative is small.
Further, as mentioned above, the lowest nonzero derivative is of $3^{rd}$} {{\color{black}order in the residue compared to $2^{nd}$ order in the conventional finite difference method.
Similar expressions for $\dfrac{\partial H}{\partial q'^{(1)}}\; \& \; \dfrac{\partial H}{\partial q'^{(2)}}$ can be obtained by putting $i=1,\;2$  respectively and writing  a Taylor series expression 
}}
{The differentiation of the generalised coordinates $(q'^{(1)},q'^{(2)},q'^{(3)})$ can be carried out as}:
\begin{eqnarray*}
 \dfrac{\partial{q'^{(i)}}}{\partial t}&=\dfrac{\partial{q'^{(i)}}}{\partial x'_j}\dfrac{d{x'_j}}{dt}=\left[ \nabla q'^i\right]_j\,2\left[\widetilde{\Lambda}\widetilde{\Lambda}^T\right]_{jl} k_l.
\end{eqnarray*}

Let $2\left[\widetilde{\Lambda}\widetilde{\Lambda}^T\right]_{jl} k_l=v_j$ where $\bm v$ is a vector whose $j^{th}$ component in
Cartesian coordinates is $v_j$, and $k_l$ are the Cartesian components of vector $\bm k$.
\begin{subequations}\label{eq: contravariant_array}
\begin{eqnarray}
 \dfrac{\partial{q'^{(1)}}}{\partial t}&=\dfrac{\partial{r'}}{\partial t}=\dfrac{\left(\bm{a}_2 \times \bm{a}_3\right)_j}{\Omega} v_j=
 \dfrac{\left(\bm{a}_2 \times \bm{a}_3\right)\cdot \bm{v}}{\Omega}\label{eq: contravariant2_1}\\
 \dfrac{\partial{q'^{(2)}}}{\partial t}&=\dfrac{\left(\bm{a}_3 \times \bm{a}_1\right)_j}{\Omega} v_j=
 \dfrac{\left(\bm{a}_3 \times \bm{a}_1\right)\cdot \bm{v}}{\Omega}\label{eq: contravariant2_2}\\
 \dfrac{\partial{q'^{(3)}}}{\partial t}&=\dfrac{\left(\bm{a}_1 \times \bm{a}_2\right)_j}{\Omega} v_j=
 \dfrac{\left(\bm{a}_1 \times \bm{a}_2\right)\cdot \bm{v}}{\Omega}\label{eq: contravariant2_3}.
\end{eqnarray}
\end{subequations}
The differential of the Hamiltonian can be written as
\begin{eqnarray*}
\dfrac{\partial H}{\partial x'_i}&=\dfrac{d{k_i}}{dt} =\dfrac{\partial H}{\partial q'^{(j)}}\dfrac{\partial q'^{(j)} }{\partial x'_i}
\end{eqnarray*}

 \begin{eqnarray}\label{eq: contravariant3}
\dfrac{d{k_i}}{dt}=\left[ 
  \begin{array}{ccc}
  \dfrac{\partial H}{\partial q'^{(j)}}  &  \dfrac{\partial H}{\partial q'^{(j)}}  & \dfrac{\partial H}{\partial q'^{(j)}}  
  \end{array}
  \right]
\left[ \begin{array}{c}
\left(\bm{a}_2 \times \bm{a}_3\right)_i \\
\left(\bm{a}_3 \times \bm{a}_1\right)_i \\
\left(\bm{a}_1 \times \bm{a}_2\right)_i
\end{array} \right]
\end{eqnarray}
To determine the path of the ray moving through the medium, the coupled differential equations, Eqs. (\ref{eq: contravariant_array})  
and  Eq. (\ref{eq: contravariant3}), have to be solved. In each of these equations, the right hand side is a function of
Cartesian components of $\bm k(k_1, k_2, k_3)$ and generalised coordinates $ (q'^1, q'^2, q'^3)$. The partial derivatives of Hamiltonian H 
at a given value of $(q'^{(1)},q'^{(2)},q'^{(3)})$ are evaluated using  Eq. (\ref{eq: numerical1}) and substituted in Eq.(\ref{eq: contravariant3}).
A program has been written in MATLAB and the differential equation toolbox has been used to solve the equations. 
The initial value of  $(q'^{(1)},q'^{(2)},q'^{(3)})$ is taken as the coordinate of the point where the incident wave touches
the outer cloak boundary. 
The initial values
for components of $\bm k$ are obtained by solving the Hamiltonian Eq. (\ref{eq: hamil2}) at the point of incidence using the following boundary
condition  on the surface \cite{pendry}:
\begin{align*}
 \bm k \times \bm n=\bm k_i \times \bm n
\end{align*}
where $\bm k_i$ is the incident wave and $\bm k$ is the wave vector inside the cloak at the point of incidence.
By solving the coupled differential equations, we  get expression of $(q'^{(1)},q'^{(2)},q'^{(3)})$  and $\bm{k}$
as a function of $t$  for the ray propagating in the cloak. For plotting the path of the ray, $(q'^{(1)},q'^{(2)},q'^{(3)})$ are 
used to calculate the position vector $\bm{x}$ in the Cartesian coordinate system.

\subsection{Piecewise linear model:} 
\label{piece_sec}
\begin{figure}[htbp]
\centerline{\includegraphics[width=8 cm]{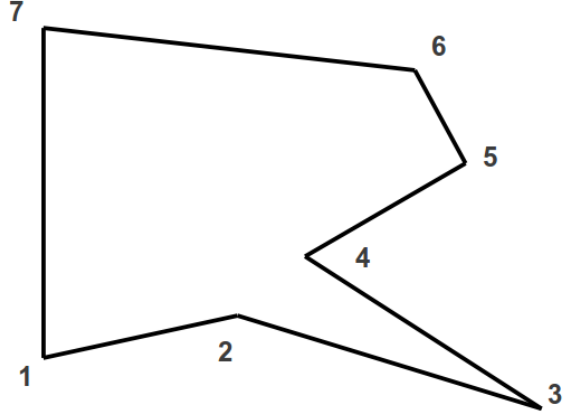}}
\caption{Piecewise linear model.}
\label{fig: piece}
\end{figure}

An example of a piecewise linear model is shown in (Fig.~\ref{fig: piece}). It is modelled as a union of  piecewise linear functions
between adjacent nodes (similar to that used in FEM). In the figure, the line joining nodes 1 and $N$ (here $N=7$) is the Z-axis.
By rotating the figure about the Z-axis, a three dimensional object is generated.


Since the body under consideration is axisymmetric, only one parameter is required to define the surface, and we denote it  as
$(q'^2=u)$.
Between $k^{th}$ and $(k+1)^{th}$ nodes, any point in the unprimed system is modelled in terms of a parameter $u$,
which takes values as $(k-1)\leq u \leq k$.
By assuming the piecewise function to be
\begin{align}\label{eqn: piecefunct}
 \bm{R}(u)&=  k \bm{r}_k- (k-1)\bm{r}_{k+1}+u(\bm{r}_{k+1}-\bm{r}_{k})\\ \nonumber&=\bm{A}u+\bm{B},
\end{align}
the terms in Eq. (\ref{eq: covariantbase}) can now be expressed as:
\begin{subequations}
\begin{align}
 \dfrac{\partial \bm{R}}{\partial u}&=\bm{A} \hspace{15mm} 
 \dfrac{\partial {R}}{\partial u}=\dfrac{\bm{R}}{R}\dfrac{\partial \bm{R}}{\partial u}=\dfrac{1}{R}\bm{R}\cdot \bm{A}\\
 \hat{\bm{R}}&=\dfrac{\bm{R}}{R}\\ 
 \dfrac{\partial \hat{\bm{R}}}{\partial u}&=\dfrac{R \dfrac{\partial \bm{R}}{\partial u}-\bm{R}\dfrac{\partial {R}}{\partial u}}{R^2}
 =\dfrac{R^2 \bm{A}-\bm{R}\left(\bm{R}\cdot \bm{A}\right)}{R^3}.
\end{align}
\end{subequations}

In order to verify the proposed algorithm, a MATLAB based simulation is carried out using $\bm{R}(q'^{(2)})$ as a piecewise function defined by Eq. (\ref{eqn: piecefunct}) 
to interpolate between successive points. In the first case, the nodal points are chosen to be  on the surface of an ellipsoid.
{{The coordinates of the position vector $\bm{r}$ are plotted inside the cloak for two of three different sets of nodes as shown
in (Fig. \ref{fig:plot_ellipsoid5}) and (Fig. \ref{fig:plot_ellipsoid10}). In (Fig. \ref{fig:plot_ellipsoid5}), the  solid  coloured lines depict the paths of the rays  to traverse half the cloak: the path in 
green is of the ray when only 5 nodes are used, the path in black  when 25 nodes are used, while red  denotes the path when 65 nodes are used.
The dashed lines in the figure, which are in red, indicate the inner and outer boundaries of the cloak for 65 nodes
while the dashed lines in green indicate the boundaries for 5 nodes. It can be observed }{{from the figure, that not only there is a prominent difference in the paths traced
by the rays for 5 nodes (solid green curve) and 65 nodes (solid red curve),  the geometry of the cloak has also been affected (observe the change in geometry from a polygon for 5 nodes to an ellipsoid for 65 nodes) causing a change in the angle of incidence.
In (Fig. \ref{fig:plot_ellipsoid10}), the  solid  coloured lines depict the paths of the rays  to traverse half the cloak: the path in 
blue is of the ray when only 10 nodes are used, the path in black  when 25 nodes are used, while red  denotes the path when 65 nodes are used.
The dashed lines in the figure, which are in red indicate the inner and outer boundaries of the cloak for 65 nodes
while the dashed lines in blue indicate the boundaries for 10 nodes. It can be observed in this case also, that there is a  difference in the paths traced
by the rays for 10 nodes (solid blue curve) and 65 nodes (solid red curve), and  the geometry of the cloak has also been affected,  but that change is less 
conspicuous than in the previous case for 5 nodes.}

{\textcolor{black}{It can be concluded that as the number of nodes is changing, the geometry gets modified because of the piecewise linear function involved,
causing a change in the point of incidence. The modelling with different sets of points is done to bring out the dependence
of the path of a ray on the number of points used in the model. To elucidate the importance of nodes required for plotting a curve,
an ellipsoid as shown (Fig. \ref{fig:plot}) is plotted for two sets of nodes: 5 and 20. For the smaller number of nodes,
the curve generated is a polygon and the point of incidence gets changed because of the distorted geometry. 
It can be observed from  (Fig. \ref{fig:plot_ellipsoid5}) and  (Fig. \ref{fig:plot_ellipsoid10}), that the smoothness and accuracy of the path traced improves as the number of nodes is increased}.
 }}\\


\begin{figure}[t!]
\centerline{\includegraphics[width=12 cm]{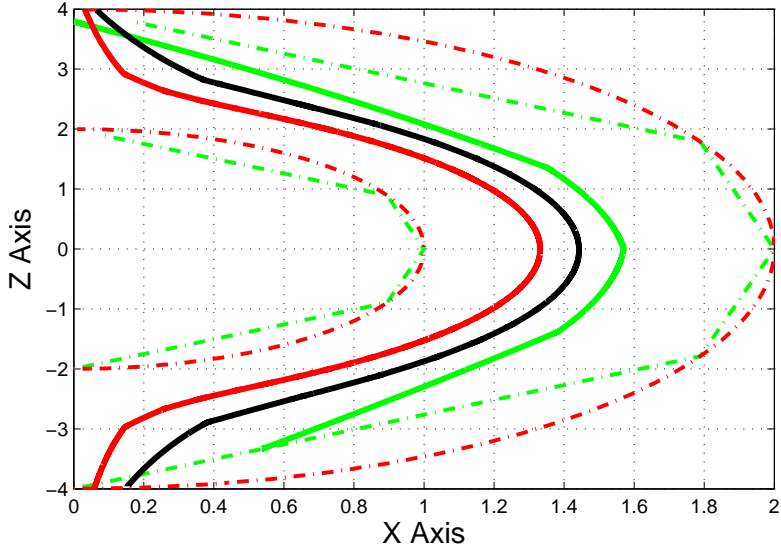}}
\caption{Two-dimensional plot of an ellipsoidal cloak  in the X-Z plane: the solid lines are the paths of the rays for different number of nodes; 
green colour for 5 nodes, black colour for 25 nodes and red for 65 nodes. The dashed lines in red indicate the inner and outer boundaries of the cloak for 65 nodes
while the dashed lines in green indicate the boundaries for 5 nodes.}
\label{fig:plot_ellipsoid5}
\end{figure}

\begin{figure}[h]
\centerline{\includegraphics[width=12 cm]{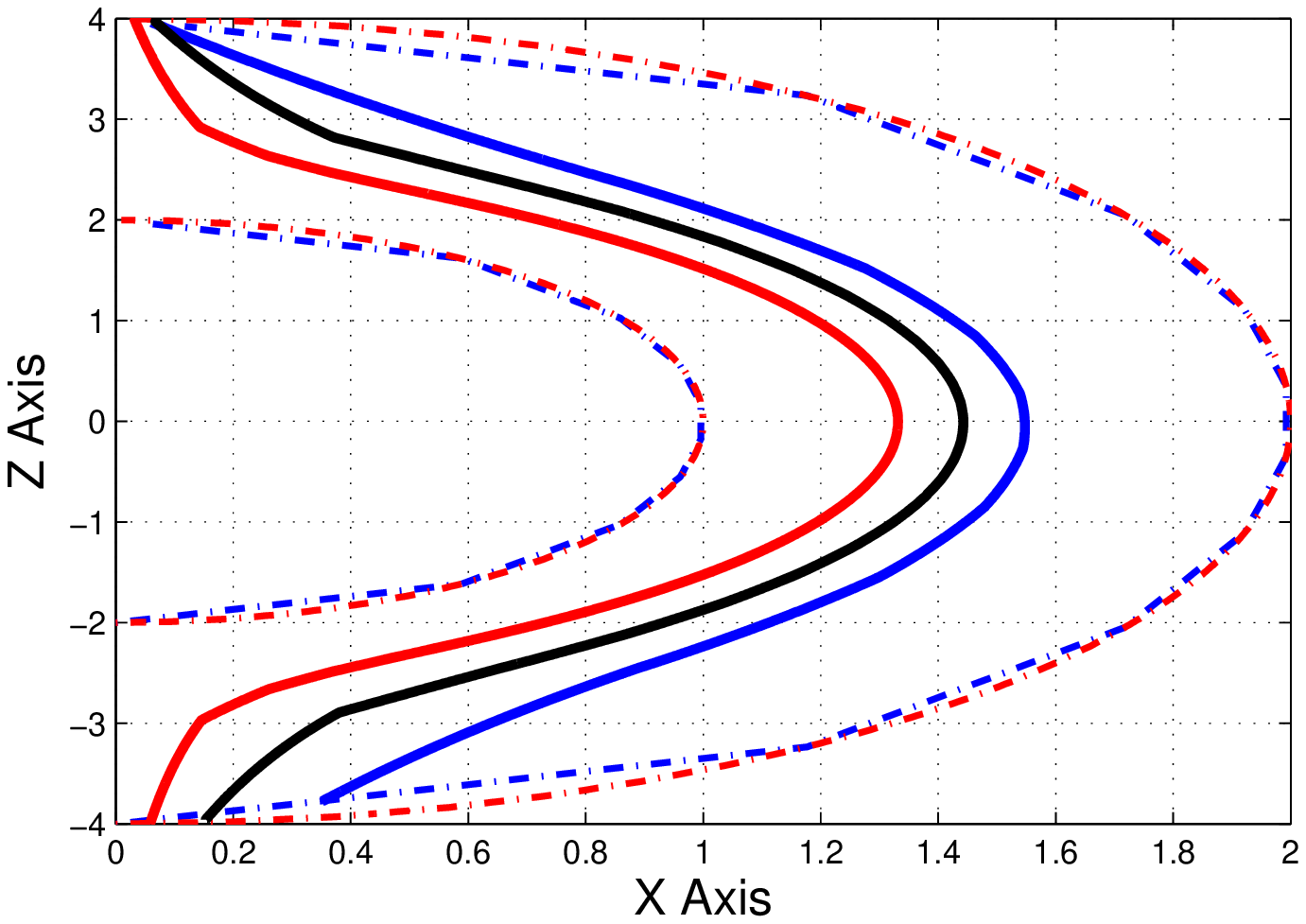}}
\caption{Two-dimensional plot of an ellipsoidal cloak  in the X-Z plane: the solid lines are the paths of the rays for different number of nodes; 
blue colour for 10 nodes, black colour for 25 nodes and red for 65 nodes. The dashed lines in red indicate the inner and outer boundaries of the cloak for 65 nodes
while the dashed lines in blue indicate the boundaries for 10 nodes.}
\label{fig:plot_ellipsoid10}
\end{figure}

\begin{figure}[ht!]
\centerline{\includegraphics[width=10 cm]{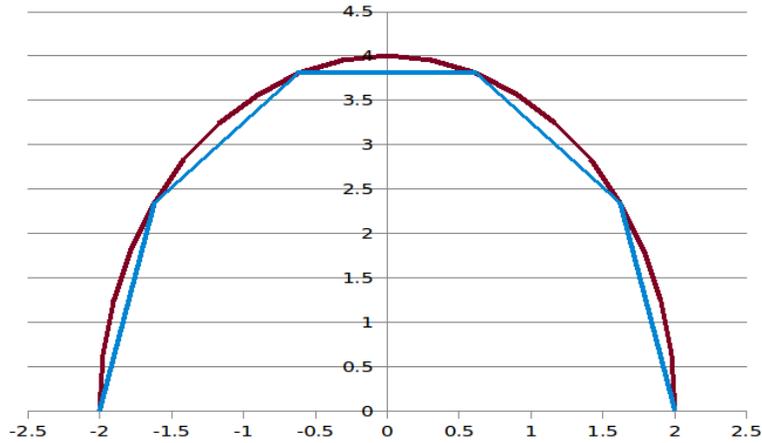}}
\caption{Plotting of an ellipsoid for 2 sets of nodes: blue coloured curve is for 5 nodes while maroon coloured curve  is for 20 nodes.}
\label{fig:plot}
\end{figure}

\begin{figure}[t!]
\centerline{\includegraphics[width=12 cm]{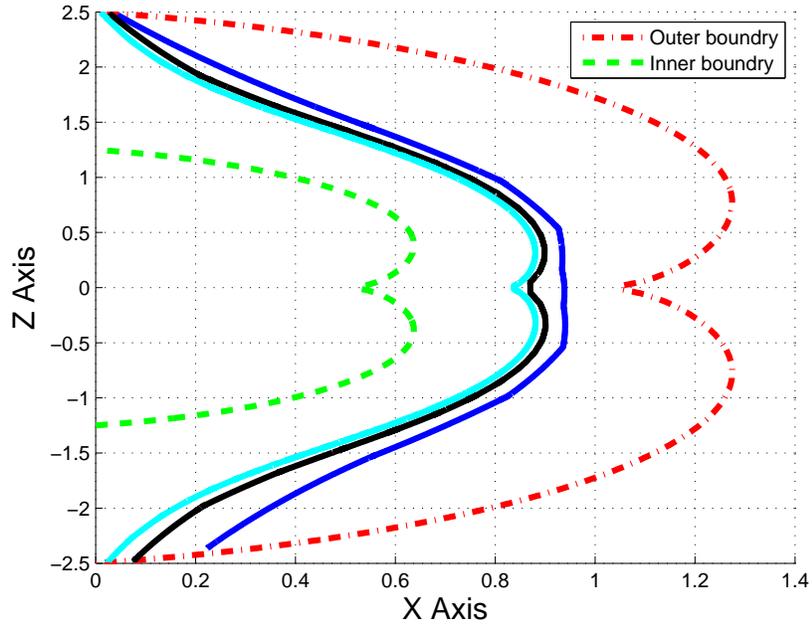}}

\caption{Two-dimensional plot of a concave surface cloak  with moderate curvature in the X-Z plane: the  blue coloured curve shows the path of the ray for 10 nodes, black coloured curve shows the path for 30 nodes, and cyan colour shows the path for 100 nodes.}
\label{fig:heartplot1}
\end{figure}

\begin{figure}[t!]
\centerline{\includegraphics[width=12 cm]{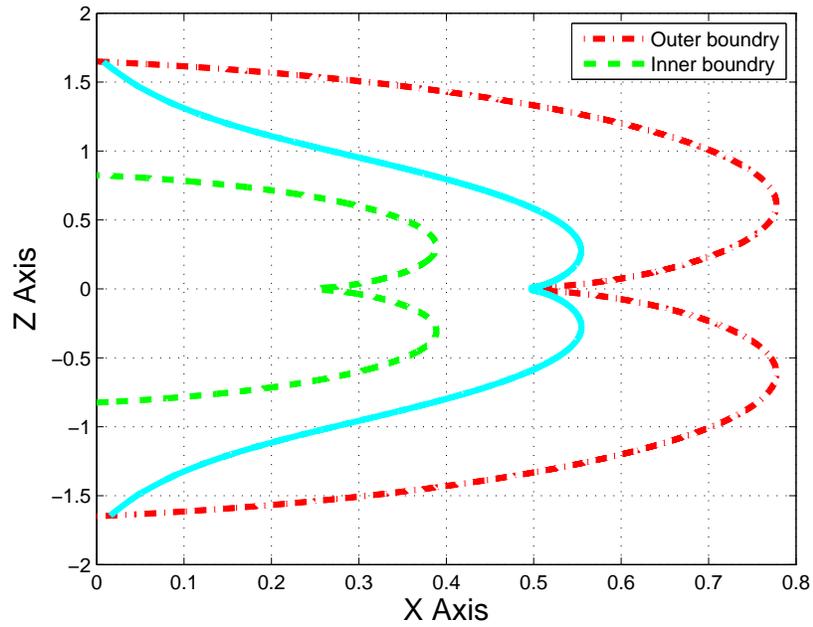}}
\caption{Two-dimensional plot of a concave surface cloak in the X-Z plane where the cyan coloured wave is the path of the ray in a cloak with a sharp inward curve.}
\label{fig:heartplot2}
\end{figure}


\begin{figure}[t!]
\centerline{\includegraphics[width=11 cm]{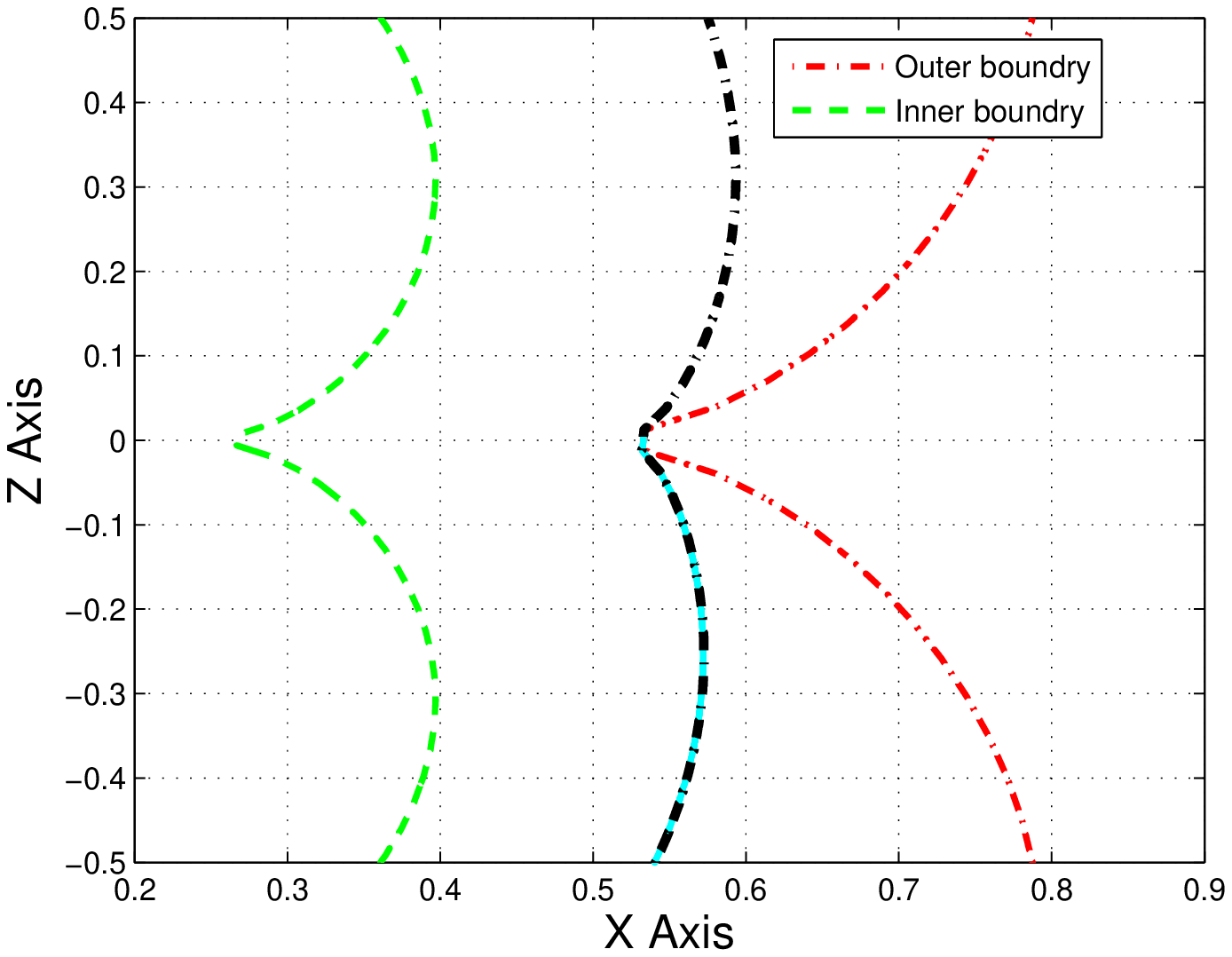}}
\caption{Comparison of paths  of the ray plotted in  a concave surface cloak in the X-Z plane using the imaginary method (Eq. (\ref{eq: numerical1})) and the conventional finite difference method where the cyan coloured wave is the path of the ray using the conventional finite difference method.
while the black coloured wave is plotted using the imaginary method.}
\label{fig:heartshapecomp}
\end{figure}


In order to verify the validity of the algorithm for concave surfaces, the nodal points are chosen to be on a concave surface and the 
MATLAB based simulation is carried out using $\bm{R}(q'^{(2)})=\bm{R}(u)$ as the piecewise function defined by Eq. (\ref{eqn: piecefunct}). 
{\textcolor{black}{The path of the ray 
is plotted inside the concave surfaced cloak with moderate curvature for three different sets of nodes as shown
in (Fig. \ref{fig:heartplot1}). It can be observed that due to the usage of only 10 nodes for the blue curve, the algorithm is not able to
capture the cusp in the geometry. As the number of nodes is increased, the variations in geometry are faithfully reproduced and the path of the ray is in accordance with
the actual geometry of the cloak. In order to test the robustness of the algorithm, the sharpness of the concave curve is increased further
and ray tracing is carried out as seen in (Fig. \ref{fig:heartplot2})}. The cyan coloured curve depicts the path of the position vector $\bm{r}$
turning just next to the sharp edge of the concave curve, and it is also within the outer boundary of the cloak as desired.
{\color{black}As the sharpness of the curve increases further, piecewise linear model might not be sufficient 
 to handle those sudden variations. (Fig. \ref{fig:heartplot2}) shows the limiting case of the usage of linear function for parameterising, where the
 ray to be traced just manages to turn sufficiently to avoid entering the body to be cloaked.
 A quadratic or even higher order function would be required to model the surface with greater sharpness.
 Further, if NURBS is used, it would be possible to model any surface with complicated geometry}. 
{\color{black} In (Fig. \ref{fig:heartplot2}), in order to compare the two numerical differentiation techniques, 
one using imaginary method  Eq. (\ref{eq: numerical1}) and the other, conventional finite difference method,the paths of the ray are tested 
on linear fitting and a parameter $du$ is varied such that there is a difference between the paths using both the numerical differentiation techniques.
The results are plotted  in  region between  $(-0.5)$ and  $(0.5)$ of the $Z$ axis to highlight the region around the cusp. It can be observed in (Fig. \ref{fig:heartshapecomp}), that
the cyan coloured ray which is plotted using conventional finite difference method hits the outer boundary of the} 
{\color{black} cloak while the black
dotted ray plotted using imaginary method differential technique manages to circumvent around the outer edge of the cloak and continue its path further}.
 
 {\color{black}The Hamiltonian H is function of the Jacobian 
  $\widetilde{\Lambda}_{il}$ in Eq. (\ref{eq: jacobfinal})and Eq. (\ref{eq: jacob2}) which depend on the parametric equation of the outer surface $\bm{R}(q^2, q^3)$ (generalised case) or $\bm{R}(q^2)$ 
  (axisymmetric case) and the derivatives of $\bm R$ with respect to the generalised coordinates. Now $\bm R$ in turn is a function of nodes, Eq. (\ref{eqn: piecefunct}), which implies 
  that the Hamiltonian is a function of the nodes selected}.

\clearpage
\subsection{General parametric representation}
{\color{black}The algorithm described in Sec. (\ref{piece_sec}) uses a piecewise linear representation of an axisymmetric surface. However,
the formulation is not limited to its parametric equation. As seen from Eq. (\ref{eq: jacobfinal}), the Hamiltonian can be calculated
if the parametric representation $\bm{R}(q'^{(2)},q'^{(3)})$ is given. A general parametric representation can be described by NURBS \cite{nurbs}.
However, the derivatives $\dfrac{\partial \bm R}{\partial q'^{2}}$ and $\dfrac{\partial \bm R}{\partial q'^{3}}$ cannot be found analytically.
Special routines need to be written to evaluate the derivatives, though they are not difficult to implement.
The other equations required for ray tracing, Eq. (\ref{eq: numerical1}), Eqs. (\ref{eq: contravariant_array}) and Eq. (\ref{eq: contravariant3}),
can all be generalised for that case. Thus any arbitrary parameterisation of a surface can be implemented}.\\
\textcolor{black}{In case of a cloak whose outer surface is not conformal with the body to be cloaked, Eq. (2) of \cite{nonconfor} can be adapted to use the 
technique described here, but that is beyond the scope of the present work}.

\subsection{Numerical stability}
{\color{black}The numerical method used for solving the Hamiltonian equation Eq. (\ref{eq: hamil2}) which is an ordinary differential equation 
linearises it. In Eq. (\ref{eq: jacobprime}), when the input wave is close to the inner surface of the cloak, the denominator
of the equation tends to zero since $(r'~\rightarrow ~\tau R)$. The relative permittivity and permeability Eq. (\ref{eq: epsilonprime}) become very large. For this case, the eigenvalues of the linearized
system differ by several orders of magnitude, or they also change during integration \cite{numerical_czech} which leads to instability in the solution.
In the method used in \cite{pendry}, the independent variables are the Cartesian coordinates used to trace the path of the ray. Usage
of this method leads to numerical instability for the case when the input wave is close to the inner surface of the cloak. 
In the method used in this paper, the independent variables are the generalised coordinates $(q'^{(1)},q'^{(2)},q'^{(3)})$ where $q'^{(1)}=r'$.
The method described here circumvents the problem 
because the generalised coordinate $r'$ continuously decreases and $q'^{(2)}$ lies on a smaller curve} {\color{black}compared to the outer boundary
and which is conformal with it. The points $\bm r_i$ and $\bm r_{i+1}$ are closer to each other as compared to the outer region i.e. the 
distance between the nodes  reduces as the wave travels towarsd the inner surface. Hence the accuracy of the calculation improves}.
\section{Conclusion}
An algorithm based on piecewise linear function has been proposed for designing an arbitrarily shaped cloak in three dimensions,  which circumvents
the problem of numerical instability when the input wave is close to the inner surface.
A procedure for calculating the path of a ray through it is described using ray tracing. 
In order to  demonstrate the 
 validity of the algorithm, it has been applied to an axisymmetric ellipsoid cloak as well as an axisymmetric concave surfaced cloak. For the purpose of differentiation
 of the Hamiltonian, a numerical technique is used instead of a standard analytical approach which becomes too cumbersome in this case. 
 The versatility of this algorithm stems from the usage of a piecewise  function   
 for plotting from one point to another in the code, enabling it to effectively cloak a body  having non-convex geometry or having sharp 
edges. Also, since the algorithm does not depend on the linear nature of the function being used, higher order functions such as
quadratic ones can be used to improve the efficiency further.

\section{Acknowledgements}
The authors gratefully acknowledge the support from the Indian Institute of Technology Bombay. The authors would  also like to thank 
Prof. R. K. Shevgaonkar for his valuable suggestions throughout
this work. The help given by Mr. Chinmay Rajhans towards the plotting of curves is also
appreciated.

%
%

\end{document}